\documentclass[onecolumn,showpacs,preprintnumbers,amsmath,amssymb]{revtex4}

\usepackage{graphicx}
\usepackage{dcolumn}
\usepackage{bm}
\usepackage{amssymb}
\usepackage{amsmath}
\usepackage{latexsym}

\begin{document}

\title{The prion-like folding behavior in aggregated proteins}

\author{Yong-Yun Ji$^1$}
\author{You-Quan Li$^1$}
\author{Jun-Wen Mao$^{1,2}$}
\author{Xiao-Wei Tang$^1$}
\address{$^1$Department of Physics, Zhejiang University, Hangzhou
310027, P.R. China.\\
$^2$Department of Physics, Huzhou Teachers college, Huzhou 313000, P.R. China.}

\date{\today}

\begin{abstract}

We investigate the folding behavior of protein sequences by
numerically studying all sequences with maximally compact lattice
model through exhaustive enumeration. We get the prion-like
behavior of protein folding. Individual proteins remaining stable
in the isolated native state may change their conformations when
they aggregate. We observe the folding properties as the
interfacial interaction strength changes, and find that the
strength must be strong enough before the propagation of the most
stable structures happens.

\end{abstract}

\pacs{87.10.+e, 87.14.Ee, 87.15.-v}

\maketitle

\section{Introduction}\label{sec:level1}

The biological function of the protein is tightly related to its
conformation. The loss of biological activity of the proteins,
such as in the case of insoluble protein plaques consisting of
amyloid fibrils in organs~\cite{1,2}, may arise from the
aggregation of misfolded protein which frequently cause various
diseases. These diseases include prion diseases, Alzheimer's
disease and Parkinson's disease~\cite{3}. For the prion diseases,
the prion protein is regarded as the origin of some
brain-attacking diseases which is known as spongiform
encephalopathies. The structures of normal form of prion protein
(PrP{$^C$}) have been obtained~\cite{4,5,6}. Both the PrP{$^C$}
and the corresponding misfolded form~(PrP{$^{Sc}$}) have identical
sequences. The only difference is in their conformations which are
considered to be responsible for the aggregation and
disease~\cite{7}. Current experiment methodologies encounter
difficulties in obtaining atomic details of seed formation and
conversion from PrP{$^C$} to PrP{$^{Sc}$}~\cite{8}. Large
challenges are still in existence for people to theoretically
investigate the mechanism of seed formation and propagation at
atomic level.

The importance of protein folding has been recognized for a long
time~\cite{9,10,11}. It is widely believed that for most single
domain proteins, the native structure is the global free-energy
minimum state, and the amino-acid sequence alone encodes
sufficient information to determine its 3-D structure~\cite{11}. A
great deal of researches have been performed to get the general
properties of protein folding and interpret the basis of
misfolding diseases. Li {\it et al.}~\cite{12} introduced the HP
lattice model and presented a meaningful interpretation on the
structure selection of nature proteins by proposing a concept of
designability. Protein refolding to an alternative form has been
observed by Harrison \emph{et al.}~\cite{13} using the lattice
model in a propagatable manner. Under conditions where the normal
native state is marginally stable or unstable, two chains refold
from monomeric minimum (the normal native state) to an alternative
multimeric minimum in energy, comprising a single refolded
conformation that can propagate itself to other protein chains.
Harrison \emph{et al.} treated both 2-D and 3-D HP models to
investigate the dimer formation and found that the structures in
the homodimeric native state re-arrange so that they are very
different in conformation from those at the monomeric native
state~\cite{14}. Giugliarelli~\emph{et al.} showed how the average
inter-amino acid interaction affects the properties of both single
and interacting proteins in a highly ordered aggregation, but only
in 2-D lattice model. They found the propensity to structural
changes of aggregated protein, namely, the prion-like behavior of
protein~\cite{15}. A Monte Carlo simulation has been used by
Bratko and Blanch to examine the competition between
intramolecular interactions which is responsible for the native
protein structure, and intermolecular association which results in
the aggregation of misfolded chains~\cite{16}. These works
enhanced our understanding on the diseases caused by protein
misfolding and aggregating. However, the folding behavior of all
sequences is mostly studied within two-dimensional models.

As we have known that the lattice model used to be a valuable
model for the designability, it is  therefore interesting to
investigate the behavior of protein misfolding and aggregating
together with the changes of designability and stability on the
basis of lattice model. In this paper we apply the 3-D HP model to
elucidate the mechanism of protein aggregation diseases such as
prion diseases. In Sec.~\ref{sec:model}, we introduced our model
that consists of a stack of twenty-seven toy bricks, each of them
stands for a $3\times3\times3$ cubic lattice. This imitates 27
aggregated proteins interacting with each other through interface.
In Sec.~\ref{sec:discuss}, we discussed the effect of interaction
between proteins on the structural stability. Our results show the
conversion of most stable structures between isolated state and
aggregated state. The change of most stable structures is strongly
correlated with the strength of interfacial interaction.

\section{``Toy Brick" Model}\label{sec:model}

Both two- and three-dimensional lattice models which describe the
protein folding into the native structure as free energy
minimization are NP-hard~\cite{17}. It is significant to introduce
simplified models to achieve some essential properties of protein
folding. The HP lattice model introduced first by Dill~\cite{18}
has helped one in understanding essential properties of protein
folding and evolution. That model and its extended ones are widely
used up to date \cite{19,20,21}, and we have investigated the
medium effects on the selection of sequences folding into stable
proteins with the simple HP model and got some meaningful
results~\cite{22}.

\begin{figure}
\includegraphics[width=0.4\textwidth]{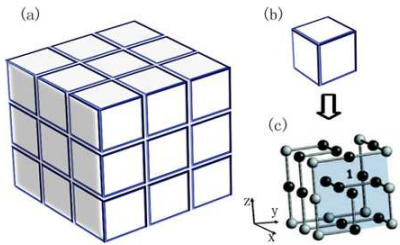}
\caption{\label{fig_1} The multimer stacked with 27 proteins (a)
and an example structure of single proteins (c) with hydrophobic
residues (grey) and polar residues (black). }
\end{figure}

In order to investigate the folding behavior of aggregated protein
(multimer), we reconstruct the original HP model by putting a
number of individual model proteins~(monomers) together, namely,
building the cubic toy bricks layer and layer in order. Each
monomer has identical sequence and structure but is in different
orientation. In this paper, we study the multimer which is stacked
by $3{\times}3{\times}3$ ordered monomers(Fig.~\ref{fig_1}~(a)).
As is shown in Fig.~\ref{fig_1}~(c), each monomer is figured as a
cube formed by a chain of 27 beads occupying the discrete sites of
a lattice in a self-avoiding way, with two types of beads of polar
(P) and hydrophobic (H) amino-acids respectively. These monomers
interact with each other through the contact amino-acids at the
surfaces of adjacent monomers respectively. In this model, the
total energy of the aggregated protein includes the pair contact
energy of the amino acids inside each monomer and the interfacial
potential ~(the additional energy) caused by the contact of amino
acids from two adjacent monomers. In our case, 54 interfaces
between
\begin{math}3{\times}3{\times}3\end{math} ordered monomers should
be considered. Thus, the energy of the multimer is given by
\begin{equation}\label{eq:hamiltonian}
H=27{\times}\sum_{i<j}E_{\sigma_i\sigma_j}\delta_{|r_i-r_j|,1}(1-\delta_{|i-j|,1})
+{\alpha}\sum_{\{f,f'\}}{\varepsilon}_{ff'}
\end{equation}
where the former part represents the total energy of 27 isolated
monomers, and i, j denote for the successive labels of residues in
a sequence, \begin{math}r_i\end{math} for the position (of the
\begin{math}i-th\end{math} residue) on the lattice sites, and
\begin{math}\sigma_i\end{math} for H (P) corresponding to hydrophobic
(polar) residue respectively. Here the delta notation is adopted,
{\it i.e.},
\begin{math}\delta_{a,b}=1\end{math} if a=b and
\begin{math}\delta_{a,b}=0\end{math} if \begin{math}a\ne b\end{math}.
As the hydrophobic force drives the protein to fold into a compact
shape with more hydrophobic residues inside as possible \cite{23},
the H-H contacts are more favorite in this model, which can be
characterized by choosing
\begin{math}E_{PP}=0\end{math}, \begin{math}E_{HP}=-1\end{math},
and \begin{math}E_{HH}=-2.3\end{math} as adopted in
Ref.~\cite{12}. The second term in Eq.~(\ref{eq:hamiltonian}) is
introduced to describe the interfacial energy caused by the
contact of model proteins. The sum will be done over all the 54
pairs of surfaces. The
\begin{math}\alpha\end{math} denotes for the strength of the
interfacial interaction and it ranges from 0.1 to 0.9. The
\begin{math}{\varepsilon}_{ff'}\end{math} is the pair
interfacial energy of two contact faces $f$ and $f'$ which belong
to two adjacent monomers respectively.

For calculation convenience, we label the six faces of model
protein with $ 1, 2, 3, \overline{1}, \overline{2}, \overline{3}$
for a given structure, of which the normal directions correspond
to the six directions along
$\textbf{x},\textbf{y},\textbf{z},\textbf{-x},\textbf{-y}$, and
$\textbf{-z}$ respectively (Fig.\ref{fig_1}(c)). For each face
there are still four states related by rotation of $\pi/2$ that
are specified by second label 0,$\pi/2$,$\pi$ and $3\pi/2$. Then
we can easily denote each face of the protein with brackets:
$|\mu,\theta\rangle$, where $\mu$ is the face label and $\theta$
is the rotation label. When the structure and sequence of the
protein are fixed, we can write down 24 pairs of kets and bras,
$|\mu,\theta\rangle$, ${\langle}\mu,\theta|$. The interfacial
energy between model proteins can be represented as:
\begin{math}
{\varepsilon}_{ff'}={\langle}f|f'{\rangle}
={\langle}\mu,\theta|{\mu}',{\theta}'{\rangle}
\end{math},
it denotes the contact energy between $\theta$-th rotated state of
$\mu$-th face of a monomer and ${\theta}'$-th rotated state of
${\mu}'$-th face of the adjacent monomer, and the $f$ stands for
the $\theta$-th rotated state of $\mu$-th face. The second term of
energy in Eq.~(\ref{eq:hamiltonian}) is then given by:
\begin{equation}
{\alpha}\sum_{\{f,f'\}}{\varepsilon}_{ff'}
={\alpha}\sum_{\{f,f'\}}{{\langle}f|f'\rangle}
={\alpha}\sum_{\{\mu,\theta,{\mu}',{\theta}'\}}
{{\langle}\mu,\theta|{\mu}',{\theta}'{\rangle}}
\end{equation}
Both ${\langle}\mu,\theta|$ and $|{\mu}',{\theta}'{\rangle}$ can
be represented by matrixes with entities being either $H$ or $P$.
In terms of these matrixes, we can easily get the contact energy
of two faces by accounting the pair contact energies of
corresponding matrix entities, namely,
\begin{equation}
{{\langle}\mu,\theta|{\mu}',{\theta}'{\rangle}}={\sum_{k=1}^3}{\sum_{l=1}^3}E_{\sigma_{M_{kl}},\sigma_{N_{kl}}}.
\end{equation}
Here $M$ and $N$ stands for the aforementioned matrixes
corresponding to ${\langle}\mu,\theta|$ and
$|{\mu}',{\theta}'{\rangle}$. To avoid the ambiguous in defining
the matrices of ${\langle}\mu,\theta|$ and $|\mu,\theta{\rangle}$,
we make the following appointment. The matrix is determined by the
face of the cubic by taking right hand and head along positive
directions of two axis respectively, Additional appoint is that we
look toward inside the cubic to define a ket while toward outside
to define a bra. For example, in $3{\times}3{\times}3$ case as
show in Fig.~\ref{fig_1}(c), the $1$-th face contact with
$\overline{1}$-th face of another monomer, we write down the
$|1,0{\rangle}$ and ${\langle}\overline{1},0|$ as:
\begin{equation}
N=|1,0{\rangle}=\left[\begin{array}{ccc}H & P & H\\H & P & H\\H &
P & H\end{array}\right],
M={\langle}{\overline{1}},0|=\left[\begin{array}{ccc}H & P & H\\P
& P & P\\H & H & H\end{array}\right].
\end{equation}
Then the contact energy of these two faces is evaluated as
\begin{eqnarray}
&&{{\langle}{\overline{1}},0|1,0\rangle}
 ={\sum_{k=1}^3}{\sum_{l=1}^3}E_{\sigma_{M_{kl}},\sigma_{N_{kl}}} \nonumber\\
&&\hspace{8mm} =E_{HH}+E_{PP}+E_{HH}+E_{PH}+E_{PP} \nonumber\\
&&\hspace{11mm} +E_{PH}+E_{HH}+E_{HP}+E_{HH}.
\end{eqnarray}
clearly, $\langle \mu,\theta | \overline{\mu'},\theta \rangle=
\langle\overline{\mu'}, \theta | \mu, \theta\rangle$ in our
definition. To minimize the total energy of the aggregated
proteins, the optimal configuration is adjusted by rotating the
bra $\langle\overline{\mu'}, \theta|$ to an appropriate
$\langle\overline{\mu'}, \theta'|$ such that the inner product
$\langle\overline{\mu'}, \theta' | \mu, \theta\rangle$ take the
smallest magnitude.

\section{Results and Discussion}\label{sec:discuss}

It has been noticed by Li \emph{et al.}~\cite{12} that some
structures can be designed by a large number of sequences, while
some ones can be designed by only few sequences. To elucidate this
difference, they introduced the designability of a structure which
is measured by the number of sequences that take this structure as
their unique lowest energy state. In conclusion, structures differ
drastically according to their designability, {\it i.e.}, highly
designable structures emerge with a number of associated sequences
much larger than the average ones. Additionally, the energy gap
represents the minimum energy a particular sequence needs to
change from its ground-state structure into an alternative compact
structure. And the average energy gap for a given structure is
evaluated by averaging the gaps over all the sequences which
design that structure. In the single-monomer HP model, the
structures with large designability have much larger average gap
than those with small designability, and there is an apparent jump
around \begin{math}N_s=1400\end{math} in the average energy gap.
This feature was first noticed by Li \emph{et al.}\cite{12}, thus
these highly designable structures are thermodynamically more
stable and possess protein-like secondary structures into which
the sequences fold faster than the other structures.

In our model, the designability and the energy gap for the
aggregated protein are similarly defined, with the energy of
single protein replaced by the total energy of the multimer.
Considering that the different orientation of the monomers
contributes differently to the total energy of the multimer in
spite of the identical sequence and structure of each monomer, the
designability (\begin{math}N_s\end{math}) in this case denotes for
the number of sequences that take the individual structure as the
unique lowest energy state of multimer. Similarly, the energy gap
\begin{math}\delta_s\end{math} is the minimum energy for the multimer
to covert from its ground-state conformation~(including 27
particularly oriented monomers) into an alternative form .

In our simulation, we search the maximal compact cubic structure
of the monomer with various designability first. Based on this,
we calculate the total energy of
the multimer by stacking 27 monomers one by one to form the
aggregated proteins. At each step the state with minimal energy
will be preserved. After all monomers are on their positions, we
regulate the orientation of each monomer again to search the
minimal energy of multimer. After the exhaustive enumeration of
all possible sequences, we find that many structures with high
average gap in isolated case are no more highly designable in
aggregated case when
\begin{math}\alpha=1\end{math}. There are 2,197,634 prion-like
sequences taking one structure as the native state(unique energy
minimum) in isolated case, but taking another structure when
aggregated. While 1,905,960 sequences take the same structures as
their native states both in isolated and aggregated cases.

\begin{figure}
\includegraphics[width=0.5\textwidth]{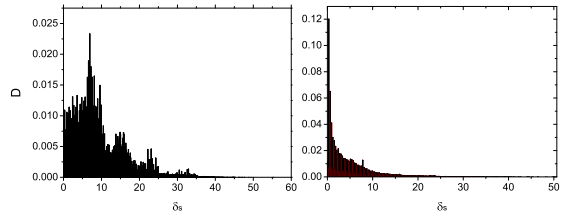}
\caption{\label{fig_2} The distribution of sequences(D) versus the
energy gap $\delta_s$ for: normal sequences(left) and prion-like
sequences(right).}
\end{figure}

By investigating the sequences of these two cases, we find that
the normal sequences possess much more proportion of sequences
with larger energy gap, though more prion-like sequences there
are. Fig.~\ref{fig_2} shows the percentage of total sequences
versus the energy gap respectively. The distribution of normal
sequences (left) has a peak at ${\delta}_s=6.9$, in the right
figure plotted for prion-like sequences, the distribution
decreases monotonously with the increase of ${\delta}_s$. In this
simplified model, the energy gap is considered to be a good
parameter to indicate the stability of a sequence at its native
state. The larger the energy gap is, the more stability the
structures have. Consequently, minority sequences are prion-like.
Actually in the nature there should be few prion-like sequences
and most protein sequences are stable in their native states. We
analyze some of the prion-like sequences by assuming that the
decrease of energy can drive the proteins in a multimer to re-fold
to new lower energy states. When the isolated native monomers
aggregate to form a multimer, some replacement of isolated native
structure by new PrP$^{Sc}$-like structure will reduce the energy
of the multimer. In this case, all monomers in the multimer will
prefer to change their structures for staying much lower total
energy level.

\begin{figure}
\includegraphics[width=0.5\textwidth]{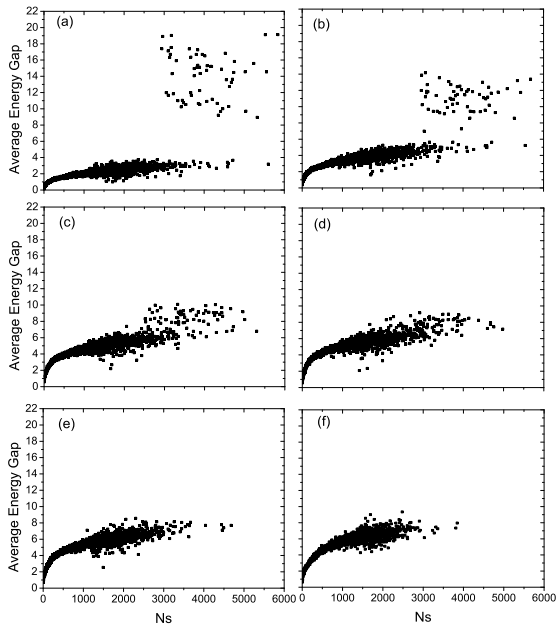}
\caption{\label{fig_3} The average energy gap versus the Ns for
different interfacial interaction strength: (a) $\alpha$=0.1, (b)
$\alpha$=0.3, (c) $\alpha$=0.5, (d) $\alpha$=0.6, (e)
$\alpha$=0.7, (f) $\alpha$=0.9.}
\end{figure}

\begin{figure}
\includegraphics[width=0.5\textwidth]{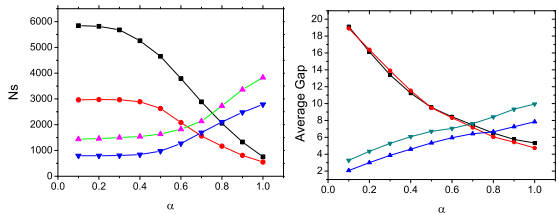}
\caption{\label{fig_4} The Ns(left) and Average Gap(right) versus
  $\alpha$ for four particular structures:  The highest designable structure
in isolated(square) and in aggregated(up triangle) case, the largest
average gap one in isolated(circle) and in aggregated(down triangle)
case.}
\end{figure}

In our simulation, we obtain that the average gaps of the
structures with high designability diminish with increasing
\begin{math}\alpha\end{math} from 0.1 to 0.9, while the average
gaps of most structures with low designability increase.
Additionally, a number of new structures which have no sequences
taking them as native state in isolated case emerge. In isolated case,
there is an abrupt jump in the average energy gap which separates
the structures into two groups, the highly designable and lowly
designable. These two groups are mixed around $\alpha=0.6$ in
aggregated case~(Fig.~\ref{fig_3}). This implies that the stable
structures in isolated case may become unstable and the new stable
structures emerge only when the strength of interaction is large
enough. This may be consistent with the result in Ref.~\cite{24}
that most misfolding diseases have a broad incubation period
before they cause symptoms and some patients will not be injured.
After the seed formation of the misfolding protein, it maybe still
take some time for misfolded proteins to get strong enough as
condition changes. If the strength stays below the transition
point, the injury will not be induced. We investigate some
particular structures. The designability and average gaps of
highest designable and largest average gap structures in isolated
case diminish continuously with increasing
\begin{math}\alpha\end{math}, but some other structures~(lowly
designable in isolated case) enhance their designability and average
gaps when aggregate~(Fig.~\ref{fig_4}).

\begin{figure}
\includegraphics[width=0.5\textwidth]{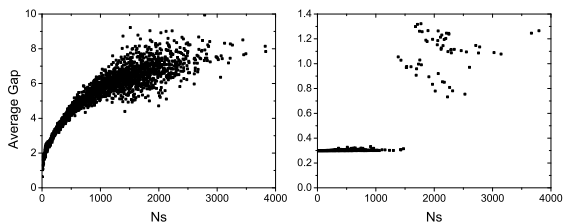}
\caption{\label{fig_5} The average energy gap versus the Ns: 27
monomers aggregated case($\alpha$=1)(left); isolated monomer
case(right). }
\end{figure}

We compare the change of average gap versus the designability in
isolated and aggregated cases respectively. In the latter case with
\begin{math}\alpha=1\end{math}, our simulation shows that the
average gap increases almost continuously with the increasing of
Ns, and there is no abrupt jump in average gap (Fig.~\ref{fig_5}).
Thus the structures can't be distinguished by the designability
and the average gap obviously. The largest average gap reaches
9.945 which is much larger than that in isolated case. Considering
an individual sequence, there are two sequences whose gaps reach
the value of 55.4 in the aggregated case~(2.6 in isolated case)
respectively. There are no sequences with energy gap larger than
70.2, the interfacial energy counterpart of these gaps. As is
shown in Ref ~\cite{12}, there are 60 highly designable structures
which are distinguished by large average gap from other ones in
isolated case. When the proteins aggregate, the average gap
increases with designability Ns continuously. But there are large
difference in the structures. We find the structures with largest
average gap or the highest designability are no more the ones in
isolated case, which implies that the most stable structures change
when the proteins aggregate. This is similar to the situation that
prion-like protein differs in their conformations, PrP$^C$ and
PrP$^{Sc}$, in isolated and aggregated cases respectively \cite{13},
which is thought to deduce the diseases.

\begin{figure}
\includegraphics[width=0.5\textwidth]{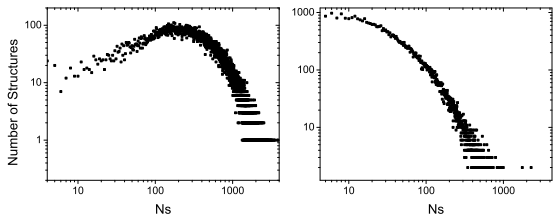}
\caption{\label{fig_6}The Ns versus the number of structures in
the situation: 27 monomers aggregated($\alpha$=1)(left); single
monomer(right).}
\end{figure}

When \begin{math}\alpha=1\end{math}, 36.7 percent of sequences
take one structure as their unique ground states, which is much
larger than that in the isolated case~(4.75$\%$). In the aggregated
case, the highest designability Ns is of 3831. Hence there must be
much more sequences which average over the other structures and
make some lowly designable structures possessing more sequences.
As is shown in Fig.~\ref{fig_6}, in aggregated~(left) case, there
are much more structures with large Ns, {\it i.e.}, there are 1497
structures whose Ns are larger than 1400. In the situation of
\begin{math}\alpha=1\end{math}, the distribution of structures
with Ns will reach a maximum when Ns=113~(Fig.~\ref{fig_6}), and
it is noticeable that the distributions are same when we change
the $\alpha$ from 0.000001 to 1. In comparison with the isolated
case, there are much less structures that take for individual
small designability : in aggregated case the largest number of
structures gets 111 at $Ns=113$, which is 1109 with Ns being 2 in
isolated case. The sequences are much more averaged in all
structures in aggregated case, but there are still some structures
which are occupied by large number of sequences, and particularly
with large average gap. The average gap changes from 0.65~
(\begin{math}Ns=1\end{math}) to
9.95~(\begin{math}Ns=2785\end{math}), and the largest
designability Ns is 3831 with average gap being 7.85.

In summary, by 3-D HP lattice model we observed that there are
some sequences which do have prion-like behavior when aggregated.
It has been known that the thermodynamic stability of proteins in
solution is affected by a variety of factors including
temperature, hydrostatic pressure, and presence of additives, such
as salts and cosolvent species, which implies that the interfacial
interactions between isolated proteins are ubiquitous. If the
concentration of protein is sufficiently high, these isolated
proteins have a tendency to aggregate reaching a lower energy
state. Our calculation showed that a few proteins will propagate
the aggregated normal form to abnormal conformation to get more
stable multimer~(with lower energy), namely, the prion-like
behavior. Particularly, we found that the most stable structures
are no more the ones in isolated case when the proteins aggregate,
such as the structures with largest average gap or the highest
designability. Furthermore, we obtained that the average gaps of
the structures with high designability diminish with increasing
\begin{math}\alpha\end{math}, while the average gaps of most
structures with low designability increase. Thus there is no
obvious jump in average gap between lowly and highly designable
structure for the aggregated proteins. We expect this result can
give some hints on the study of misfolding diseases. Since this is
a simplified model, further understandings about protein
aggregation and misfolding diseases are expected. A more realistic
protein-like model with the various interaction being close to the
real proteins is in progress.

\begin{acknowledgments}
This work is supported by NSFC No.10225419 and the Key Project of
Chinese Ministry of Education No.02046.
\end{acknowledgments}


\end{document}